\def\Title{Measurements of Nondegenerate Discrete Observables}
\def\Author{Masanao Ozawa}
  \documentstyle[preprint,aps,mynum]{revtex}
  \tightenlines                             
  \newcommand{\beq}{\begin{equation}}
  \newcommand{\eeq}{\end{equation}}
  \newcommand{\beql}[1]{\begin{equation}\label{eq:#1}}
  \newcommand{\beqa}{\begin{eqnarray}}
  \newcommand{\eeqa}{\end{eqnarray}}
  \newcommand{\beqas}{\begin{eqnarray*}}
  \newcommand{\eeqas}{\end{eqnarray*}}
  \newtheorem{Theorem}{Theorem}
  \newcommand{\R}{{\bf R}}
  \newcommand{\bA}{{\bf A}}

  \newcommand{\bL}{{\bf L}}
  \newcommand{\bM}{{\bf M}}

  \newcommand{\bP}{{\bf P}}
  
  \newcommand{\bS}{{\bf S}}
  \newcommand{\bT}{{\bf T}}
  
  \newcommand{\bW}{{\bf W}}
  \newcommand{\bX}{{\bf X}}
  \newcommand{\bY}{{\bf Y}}

  \newcommand{\cD}{{\cal D}}
  \newcommand{\cE}{{\cal E}}

  \newcommand{\cH}{{\cal H}}

  \newcommand{\cK}{{\cal K}}
  \newcommand{\cL}{{\cal L}}

  \newcommand{\al}{\alpha}
  \newcommand{\be}{\beta}
  
  \newcommand{\da}{\dagger}

  \newcommand{\et}{\eta}

  \newcommand{\la}{\lambda}
  \newcommand{\mb}{\mbox}
  \newcommand{\nn}{\nonumber}
  
  \newcommand{\ph}{\phi}
  \newcommand{\ps}{\psi}
  \newcommand{\rh}{\rho}
  \newcommand{\si}{\sigma}
  \newcommand{\ta}{\tau}

  \renewcommand{\vr}{\varrho}
  
  \newcommand{\tc}{\tau c}
  \newcommand{\De}{\Delta}
  \newcommand{\Ph}{\Phi}
  \newcommand{\Ps}{\Psi}
  \newcommand{\Tr}{\mbox{\rm Tr}}
  \newcommand{\ba}{{\bf a}}
  \newcommand{\bb}{{\bf b}}

  \newcommand{\bx}{{\bf x}}
  \newcommand{\by}{{\bf y}}
  \newcommand{\eq}[1]{(\ref{eq:#1})}
\newcommand{\bold}[1]{\mbox{\boldmath{$#1$}}}
\newcommand{\bra}[1]{\langle#1|}
\newcommand{\ket}[1]{|#1\rangle}

\newcommand{\bracket}[1]{\langle#1\rangle}
\newcommand{\bvr}{\mbox{\boldmath{$\vr$}}}
\title{\bf \Title}
\author{\Author}
\address{School of Informatics and Sciences,
Nagoya University, Nagoya 464-8601, Japan}
\begin{document}
% TITLE & ABSTRACT:
\draft
\preprint{}
\maketitle
\begin{abstract}
Every measurement on a quantum system causes a state change
from the system state just before the measurement to the system state 
just after the measurement conditional upon the outcome of measurement.
This paper determines all the possible conditional state changes
caused by measurements of nondegenerate discrete observables.
For this purpose, the following conditions are shown to be equivalent
for measurements of nondegenerate discrete observables: (i) The joint 
probability distribution of the outcomes of successive measurements depends 
affinely on the initial state. (ii) The apparatus has an indirect 
measurement model. (iii) The state change is described by a positive 
superoperator valued measure. (iv) The state change is described by 
a completely positive superoperator valued measure.  
(v) The output state is independent of 
the input state and the family of output states can be arbitrarily 
chosen by the choice of the apparatus.
The implications to the measurement problem are discussed briefly.
\end{abstract}
% TEXT:
\narrowtext
\section{Introduction}\label{se:1}
\setcounter{equation}{0}

Every measurement on a quantum system causes a quantum state reduction,
a state transformation $\rh\mapsto\rh_{x}$ 
from the system state $\rh$ just before the measurement
to the system state $\rh_{x}$
just after the measurement conditional upon the outcome $x$
of measurement.
In order to determine the quantum state reduction caused by a 
measurement of a given observable, von Neumann posed the 
{\em repeatability hypothesis} \cite[p.~335]{vN55}: 
If an observable is measured twice in succession 
in a system, then we get the same value each time.
Then, a measurement of a discrete observable $A$ satisfies the 
repeatability hypothesis if and only if $\rh_{x}$ is a mixture of
eigenstates of $A$ corresponding to the eigenvalue $x$.
Thus, under this hypothesis
the measurement of a nondegenerate discrete observable $A$
causes the unique quantum state reduction such that $\rh_{x}$ is
the unique eigenstate corresponding to the eigenvalue $x$.
However, von Neumann admitted that there are many quantum state
reductions caused by measuring the degenerate observable $A$ 
even when the repeatability hypothesis holds \cite[p.~348]{vN55}.
Thus, for degenerate observables further hypothesis were demanded.
In order to characterize the least disturbing measurement,
L\"{u}ders \cite{Lud51} posed the {\em projection postulate}:  
The measurement of a discrete observable $A$ in the state $\rh$
leaves the system in the state $\rh_{x}=E_{x}\rh E_{x}/\Tr[E_{x}\rh]$, 
where $E_{x}$ is the eigenprojection corresponding to the eigenvalue $x$.
Obviously, the projection postulate implies the repeatability
hypothesis and determines the output state uniquely,
even for the degenerate discrete observable $A$.

Despite the above attempts, Davies and Lewis \cite{DL70} conjectured
that no measurements of continuous observables satisfy the repeatability 
hypotheses and proposed abandoning the repeatability hypothesis.
Actually, their conjecture was proved later in \cite{84QC,85CA};
the essential part of the proof given in \cite{85CA} shows that 
even in the measurement of a continuous observable the output state 
conditional upon the outcome can be still described by a density operator.
Moreover, it can be readily seen that there are many ways of measuring 
the same observable without satisfying the repeatability hypothesis
such as photon counting \cite{IUO90}, which arises every optical 
experiment, and contractive state measurement 
\cite{Yue83,88MS,Mad88,89RS}, which beats the standard quantum 
limit for monitoring the free-mass position claimed 
in \cite{BV74,CTDSZ80,Cav85}.

Once we abandon the repeatability hypothesis or the projection
postulate, the problem of determining all the possible quantum
state reductions caused by measurements of a given observable
has a primary importance in quantum mechanics.
Especially, the problem receives increasing interests recently
not only from the foundational point of view but also from the technological 
point of view, since the measurement is used for preparing the state of 
the system in such  processes as purification procedures 
in the field of quantum information \cite{BDSW96,VPRK97}.

The purpose of this paper is to give the complete solution to the 
above problem for the measurements of nondegenerate discrete
observables.
It will be shown that a surprisingly general condition for the 
measurement statistics suffices to determine all the possible
quantum state reductions realized by indirect measurement models.
It is shown that for measurements of nondegenerate discrete
observables the output state is independent of the input state 
in any measurement and that the family of output states can be 
arbitrarily chosen by the choice of the apparatus.
Moreover, all of them are shown to have indirect measurement
models.

In order to obtain a mathematical description of quantum state 
reductions for the most general class of measurements 
we consider the two requirements:
one is necessary and the other is sufficient.

The necessary one is the mixing law of the joint probability
that requires that the joint probability distribution of 
the outcomes of the successive measurement depends affinely 
on the input state.
We require this condition as a necessary condition for every 
apparatus to satisfy.
It will be shown that this is equivalent to the requirement 
that every apparatus has a normalized positive superoperator 
valued measure that satisfies the Davies-Lewis description of 
conditional state transformations \cite{DL70,Dav76}.
The notion of normalized positive superoperator valued measures 
was first introduced by Davies and Lewis \cite{DL70} to obtain a general 
description of conditional state transformation by unifying 
the notions of operations \cite{HK64}, effects \cite{Lud67}, 
and probability operator valued measures \cite{Hel76,Hol82}.
Thus, the problem of determining possible quantum state reductions 
is reduced to the problem as to which normalized positive 
superoperator valued measure corresponds to an apparatus.

The sufficient condition is the unitary realizability condition that
requires the existence of an indirect measurement model
comprising of the probe preparation, the measuring interaction
with unitary time evolution, and the probe detection.
We require this condition as a sufficient condition so that if
a normalized positive superoperator valued measure has an indirect 
measurement model then the corresponding apparatus exists.
It was proved in \cite{83CR,84QC} that this condition is equivalent 
to the condition that the normalized positive superoperator valued 
measure is completely positive.

According to the above approach,
the class of possible quantum state reductions is included in the class 
of conditional state transformations satisfying the mixing law, i.e., 
the normalized positive superoperator valued measures, 
and includes the one satisfying the realizability condition, 
i.e., the normalized completely positive superoperator valued measures.
These two classes are generally different.

Nevertheless, for the case where $A$ is nondegenerate,
this paper shows, the above two conditions are actually equivalent.
Thus, both of them are necessary and sufficient and we reach a 
clear-cut conclusion.
According to the analysis developed in this paper, 
for any apparatus $\bA$ measuring a nondegenerate discrete observable 
$A=\sum_{n}a_{n}\ket{\ph_{n}}\bra{\ph_{n}}$ there is a sequence
$\{\bvr_{n}\}$ of density operators independent of the input 
state $\rh$ such that the measurement leaves the system
in the state $\bvr_{n}$ with the probability 
$\bracket{\ph_{n}|\rh|\ph_{n}}$, 
and conversely for any sequence $\{\bvr_{n}\}$ of density operators
such an apparatus exists.

\section{Measuring Apparatuses}

Let us consider the conventional quantum-mechanical description
of the measurement of an observable represented by a self-adjoint
operator $A$ with purely discrete spectrum on a separable
Hilbert space $\cH$.
For any real number $x$ we shall denote by $E^{A}(x)$ the
projection of $\cH$ onto the subspace $\{\ps\in\cH|\ A\ps=x\ps\}$.
If $A$ has eigenvalues $a_{1},a_{2},\ldots$ then $E^{A}(a_{n})$
is the spectral projection corresponding to $a_{n}$ and
$E^{A}(x)=0$ if $x$ is not an eigenvalue of $A$.
If the state of the system at the instant before the measurement  
is given by the density operator $\rh$ on $\cH$,
then the measurement yields the outcome $a_{n}$ with the
probability $\Tr[E^{A}(a_{n})\rh]$.  
If this measurement satisfies the projection postulate \cite{Lud51},
then the state at the instant after the measurement is 
\beql{PP}
\rh_{n}=\frac{E^{A}(a_{n})\rh E^{A}(a_{n})}{\Tr[E^{A}(a_{n})\rh]}
\eeq
provided that the measurement leads to the outcome $a_{n}$.

As it can be seen from the above description, every measuring
apparatus $\bA$ has the {\em output variable} $\bx$ that takes
the outcome in each measurement carried out by $\bA$.
Thus, the output variable is a random variable
the probability distribution of which depends only on
the {\em input state}, the state of the system at the instant
just before the measurement.
Throughout this paper, we assume that {\em the output variable 
takes the values in a countable subset of the real line $\R$}.
The probability distribution 
$\Pr\{\bx=x\|\rh\}$ 
of $\bx$ in the input state $\rh$ is called the {\em output 
distribution} of $\bA$.  
The change from the unknown input state to the output
distribution is called the {\em objective state reduction}. 
Depending on the input state $\rh$
and the outcome $\bx=x$, the state $\rh_{\{\bx=x\}}$ 
just after the measurement is determined uniquely.
The state $\rh_{\{\bx=x\}}$ is called the {\em output state}
relative to the input state $\rh$ and the outcome $\bx=x$.
If the output probability of $\bx=x$ is 0, the output state 
$\rh_{\{\bx=x\}}$ is taken to be indefinite.
The change from the unknown input state to the output state 
is called the {\em quantum state reduction}.
The above two mathematical objects,
the objective state reduction and 
the quantum state reduction, are called the 
{\em statistical property} of $\bA$.
Two apparatuses are called {\em statistically equivalent} if 
they have the same statistical property.
In what follows, every apparatus is supposed to have its own 
distinctive output variable and we denote by $\bA(\bx)$ the 
apparatus having the output variable $\bx$.

In the above measurement of $A$ satisfying the projection postulate, 
let us denote the measuring apparatus by $\bA(\ba)$
where $\ba$ stands for the output variable.  
The statistical property of $\bA(\ba)$ is represented as follows.
\beqa
\mb{output distribution: }
& &\Pr\{\ba=x\|\rh\}=\Tr[E^{A}(x)\rh]\label{eq:a}\\
\mb{output state: }
& &\rh_{\{\ba=a_{n}\}}
=\frac{E^{A}(a_{n})\rh E^{A}(a_{n})}{\Tr[E^{A}(a_{n})\rh]}\label{eq:b}
\eeqa
In the above, $a_{n}$ is an eigenvalue such that $\Tr[E^{A}(a_{n})\rh]>0$.

Now, the following problem arises:
Does every measuring apparatus for the observable $A$ necessarily
have the above statistical property?
It is postulated by the Born statistical formula that
the output distribution of the measurement of the observable
$A$ satisfies \eq{a}.
Hence, every measuring apparatus for the observable $A$ satisfies
\eq{a} by definition.
The following argument will show, however, that the existence of
an apparatus satisfying the projection postulate implies 
the existence of another apparatus which does not satisfy 
the projection postulate.
Therefore, we cannot postulate that every measurement satisfies 
the projection postulate.

Suppose that the observable $Y$ has degenerate eigenvalues and
can be represented by
\beql{3}
Y=\sum_{n,m}y_{n}|{n,m}\>\langle {n,m}|
\eeq
for some orthonormal basis $\{\ket{n,m}\}$.
Consider the following process of measuring $Y$:
(i) One measures the nondegenerate discrete observable 
\beql{4}
X=\sum_{n,m}x_{n,m}|{n,m}\>\langle {n,m}|
\eeq
where $x_{n,m}$ are all different.
(ii) If the outcome $\bx$ of the $X$ measurement leads to the value
$x_{n,m}$ then the outcome $\by$ of the $Y$ measurement is determined as
$y_{n}$. 
Then, even if the $X$ measurement satisfies the projection
postulate, the $Y$ measurement does not satisfy it. 
In fact, with the probability $\bracket{{n,m}|\rh|{n,m}}$
the $X$ measurement leads to the outcome $x_{n,m}$ and leaves
the system in the state $\ket{{n,m}}\bra{{n,m}}$ by
the projection postulate.
It follows that if the outcome is $y_{n}$ then the state 
at the instant after the $Y$ measurement is given by
\beqa
\rh_{\{\by=y_{n}\}}=
\frac
{\sum_{m}\bracket{{n,m}|\rh|{n,m}}\,\ket{n,m}\bra{n,m}}
{\sum_{m}\bracket{{n,m}|\rh|{n,m}}}.
\label{eq:5}
\eeqa
The above state depends on the choice of the orthonormal basis
$\{\ket{n,m}\}$.
If $Y$ is degenerate, there are infinitely many essentially different
choices of $\{\ket{n,m}\}$ and each choice gives a process of $Y$
measurement which does not satisfy the projection postulate.

Generalizing the above, if two observables $X,Y$ has the 
relation $Y=f(X)$, then for any apparatus $\bA(\bx)$
measuring $X$
we have the apparatus $\bA(f(\bx))$ measuring $Y$
 that outputs the outcome $f(\bx)=f(x)$ whenever $\bA(\bx)$ outputs 
the outcome
$\bx=x$.
In this case, even if $\bA(\bx)$ satisfies the projection postulate,
$\bA(f(\bx))$ does not necessarily satisfies the projection postulate.
Therefore, the output distribution of $Y$ measurement is unique but
the quantum states reduction depends on the way of measuring 
the same observable $Y$. 
More general construction of measuring apparatuses that do not satisfy 
the projection postulate will be discussed in Section \ref{se:measuring}.

Can one determine all the possible quantum state reductions
arising in measuring $A$ that are allowed by
the basic principles of quantum mechanics?
This problem will be considered in the following sections.

\section{Successive measurements}

In order to clarify the operational meaning of the quantum state
reduction, we shall generalize von Neumann's idea on repeated 
measurements of the same observable \cite[pp.~211--223]{vN55}
to arbitrary pair of measuring apparatuses and consider the 
joint probability distribution of the outcomes of the two 
measurements carried out in succession.

We suppose that the $A$ measurement described in the preceding section
is immediately followed by a measurement of a discrete 
observable $B$ with eigenvalues $b_{m}$.
Then, the conditional probability of obtaining the outcome $b_{m}$ at the
$B$ measurement is $\Tr[E^{B}(b_{m})\rh_{n}]$ conditional
upon having obtained $a_{n}$ at the $A$-measurement.
From \eq{PP}, the joint probability of obtaining $a_{n}$ 
at the $A$ measurement and $b_{m}$ at the $B$ measurement is therefore
\beqa
p_{n,m}&=&\Tr[E^{B}(b_{m})\rh_{n}]\Tr[E^{A}(a_{n})\rh]\nn\\
&=&\Tr[E^{B}(b_{m})E^{A}(a_{n})\rh E^{A}(a_{n})].\label{eq:joint-AB}
\eeqa

Generally speaking, if a measurement by the apparatus $\bA(\bx)$ 
in the input state $\rh$ is immediately followed by a measurement 
by the apparatus $\bA(\by)$, the joint probability distribution 
$\Pr\{\bx=x,\by=y\|\rh\}$ of the output variables $\bx$ and $\by$
depends only on the input state $\rh$ of the first measurement and
is given by
\beql{c}
\Pr\{\bx=x,\by=y\|\rh\}
=\Pr\{\by=y\|\rh_{\{\bx=x\}}\}\Pr\{\bx=x\|\rh\}.
\eeq
This joint probability distribution has the following significant
property.
\vskip\topsep

{\bf Mixing law of the joint probability: }
{\em For any measuring apparatuses $\bA(\bx)$ and $\bA(\by)$, 
if the input state $\rh$ is the mixture of $\rh_{1}$ and $\rh_{2}$
such that $\rh=\al\rh_{1}+(1-\al)\rh_{2}$ with $0<\al<1$
then the joint probability distribution of the outcomes of the 
successive measurement satisfies}
\beqa
\Pr\{\bx\!=\!x,\by\!=\!y\|\rh\}
&=&        \al\Pr\{\bx\!=\!x,\by\!=\!y\|\rh_{1}\}\nn\\
& &\mb{}+\!(1\!-\!\al)\Pr\{\bx\!=\!x,\by\!=\!y\|\rh_{2}\}.\label{eq:c'}
\eeqa

This is justified as follows.
If the system is in the state $\rh_{1}$ with the probability
$\al$ and in the state $\rh_{2}$ with the probability $1-\al$ then
the joint probability is their mixture in the right hand side. 
On the other hand, in this case the state of the system is described 
by the density operator $\rh$ and hence the above equality holds.

In the previous example, if the observable $B$ is measured by 
an apparatus $\bA(\bb)$ then from \eq{joint-AB} we have
\beql{d}
\Pr\{\ba=a_{n},\bb=b_{m}\|\rh\}
=\Tr[E^{B}(b_{m})E^{A}(a_{n})\rh E^{A}(a_{n})].
\eeq
Obviously, this joint probability satisfies the above mixing law.

In what follows, we require the mixing law of the joint probability.
For an arbitrary apparatus $\bA(\bx)$ with the output distribution 
$\Pr\{\bx=x\|\rh\}$ and the output state $\rh_{\{\bx=x\}}$, we define
the {\em output operator} $\bX(x,\rh)$ by
\beql{f}
\bX(x,\rh)=\Pr\{\bx=x\|\rh\}\rh_{\{\bx=x\}}.
\eeq
Then, $\bX(x,\rh)$ is a trace class operator \cite{Sch60a}
determined by the statistical
property of the apparatus $\bA(\bx)$, the input state $\rh$, and
the outcome $\bx=x$.

For the measuring apparatus $\bA(\ba)$, the output operator 
$\bX_{\ba}(x,\rh)$ is given by
$$
\bX_{\ba}(x,\rh)=E^{A}(x)\rh E^{A}(x).
$$ 
The above expression extends the definition of
$\bX_{\ba}(x,\rh)$ to arbitrary trace class operators $\rh$.
Then, it is easy to see that $\bX_{\ba}(x,\rh)$ has the following 
properties: (i) $\bX_{\ba}(x,\rh)$ is a positive operator if $\rh$ is
positive, (ii) the correspondence $\rh\mapsto\bX_{\ba}(x,\rh)$ is
linear, (iii)  for any $\rh$ we have
$$
\Tr[\sum_{x}\bX_{\ba}(x,\rh)]=\Tr[\rh].
$$  
In the following, we shall show that the output operator $\bX(x,\rh)$ 
of every apparatus $\bA(\bx)$ has the above properties.

Return to the joint probability distribution $\Pr\{\bx=x,\by=y\|\rh\}$.
If one measures the observable $B$ by the apparatus $\bA(\bb)$
instead of $\bA(\by)$, from \eq{c} and \eq{f} we have 
\beqa
\Pr\{\bx=x,\bb=b_{m}\|\rh\}
&=&\Tr[E^{B}(b_{m})\rh_{\{\bx=x\}}]\Pr\{\bx=x\|\rh\}\nn\\
&=&\Tr[E^{B}(b_{m})\bX(x,\rh)].\label{eq:ob-joint}
\eeqa
Suppose that $\rh$ is the mixture $\rh=\al\rh_{1}+(1-\al)\rh_{2}$.
From \eq{c'} we have
\beqas
\lefteqn{\Tr[E^{B}(b_{m})\bX(x,\rh)]}\\
&=&
\al\Tr[E^{B}(b_{m})\bX(x,\rh_{1})]+(1-\al)\Tr[E^{B}(b_{m})\bX(x,\rh_{2})]\\
&=&
\Tr[E^{B}(b_{m})[\al\bX(x,\rh_{1})+(1-\al)\bX(x,\rh_{2})]].
\eeqas
Since $B$ is arbitrary, we have
\beql{g}
\bX(x,\rh)=\al\bX(x,\rh_{1})+(1-\al)\bX(x,\rh_{2}).
\eeq

In what follows, for any $x$ let $\bX(x)$ be the mapping that
maps a density operator $\rh$ to the trace class operator $\bX(x,\rh)$.
Since every trace class operator $\si$ can be represented as
the linear combination 
\beq\label{eq:T7}
\si=\la_{1}\si_{1}-\la_{2}\si_{2}+i\la_{3}\si_{3}-i\la_{4}\si_{4}
\eeq
with four density operators 
$\si_{1},\ldots,\si_{4}$ and four positive numbers $\la_{1},
\ldots,\la_{4}$,
we can extend the mapping $\bX(x)$ to a linear transformation 
on the space $\tc(\cH)$ of trace class operators on $\cH$
by 
\beqa\label{eq:T8}
\bX(x)\si
&=&       \la_{1}\bX(x)\si_{1}- \la_{2}\bX(x)\si_{2}\nn\\
& &\mb{}+i\la_{3}\bX(x)\si_{3}-i\la_{4}\bX(x)\si_{4}.
\eeqa
Since the decomposition $\eq{T7}$ is not unique, in order for
the extension \eq{T8} to be well-defined we need to show that
the left hand side of \eq{T8} is uniquely determined independent
of the decomposition of $\si$.  This can be proved from \eq{g} and
the proof will be shown in Appendix \ref{se:A}.

We have, therefore, shown that for every apparatus $\bA(\bx)$ 
there exists a family $\{\bX(x)|\ x\in\R\}$ of linear transformations 
on $\tc(\cH)$ such that for every density operator $\rh$, we have
\beql{h}
\bX(x)\rh=\Pr\{\bx=x\|\rh\}\rh_{\{\bx=x\}}.
\eeq
The linear transformation $\bX(x)$ defined above is called 
the {\em operation} of the apparatus $\bA(\bx)$ for the outcome 
$\bx=x$.  The family $\{\bX(x)|\ x\in\R\}$ is called the 
{\em operational distribution} of the apparatus $\bA(\bx)$.
It is obvious from \eq{h} that by taking advantage of the operational
distribution, the output distribution is represented by
\beq
\Pr\{\bx=x\|\rh\}=\Tr[\bX(x)\rh]\label{eq:0117a}
\eeq
and the output state by
\beq
\rh_{\{\bx=x\}}
=\frac{\bX(x)\rh}{\Tr[\bX(x)\rh]},\label{eq:0117b}
\eeq
where the outcome $\bx=x$ is supposed to have positive probability.

\section{Operational distributions}

In order to explore mathematical properties of the operational
distribution $\{\bX(x)|\ x\in\R\}$ of the apparatus $\bA(\bx)$,
we shall provide relevant mathematical terminology.
A linear transformation $\bL$ on the space $\tc(\cH)$ of trace
class operators on $\cH$ is said to be {\em bounded} if there is 
a constant $K>0$ such that
$$
\|\bL\rh\|_{tr}\le K\|\rh\|_{tr}
$$
for all $\rh\in\tc(\cH)$, where $\|\cdot\|_{tr}$ stands for the trace norm.
Then, the norm of $\bL$ is defined by
\beq
\|\bL\|_{tr}=\sup_{\|\rh\|_{tr}\le 1}\|\bL\rh\|_{tr}.
\eeq
A linear transformation $\bM$ on the space $\cL(\cH)$ of bounded 
operators on $\cH$ is said to be {\em bounded} if there is 
a constant $K>0$ such that
$$
\|\bM A \|\le K\|A  \|
$$
for all $A  \in\cL(\cH)$, where $\|\cdot\|$ stands for the operator
norm.
Then, the norm of $\bM$ is defined by
\beq
\|\bM\|=\sup_{\|A\|\le 1}\|\bM A\|.
\eeq

A bounded linear transformation on $\tc(\cH)$ is called a 
{\em superoperator}.
For any superoperator $\bL$ on $\tc(\cH)$,
its {\em dual superoperator} $\bL^{*}$ is the bounded linear 
transformation on $\cL(\cH)$ defined by
\beql{O5.5}
\Tr[A(\bL\rh)]=\Tr[(\bL^{*}A)\rh]
\eeq
for all $A\in\cL(\cH)$ and $\rh\in\tc(\cH)$.
In this case, we have $\|\bL\|_{tr}=\|\bL^{*}\|$.
A superoperator or a dual superoperator is said to be {\em positive}
iff it maps positive operators to positive operators.
Then, a superoperator $\bL$ is positive if and only if
so is its dual.  
A super operator or a dual superoperator is said to be
{\em contractive} iff it has the norm less than or equal to one.
Then, a superoperator $\bL$ is a contractive if and only if
so is its dual.  
We have the following characterizations of positive contractive
superoperators \cite[p.~216]{BR79}, \cite[p.~18]{Dav76}.

\begin{Theorem}\label{th:0306a}
For a positive superoperator $\bL$ the following 
conditions are all equivalent:

{\rm (i)} $\bL$ is a contractive superoperator.

{\rm (ii)} $\bL^{*}$ is a contractive dual superoperator.

{\rm (iii)} $0\le\Tr[\bL\rh]\le 1$ for all density operators $\rh$.

{\rm (vi)} $0\le \bL^{*}(I)\le I.$ 

Moreover, a superoperator $\bL$ is trace preserving, i.e., 
$$
\Tr[\bL(\rh)]=\Tr[\rh]
$$ 
for all $\rh\in\tc(\cH)$ if and only if\/
$\bL^{*}$ is unital, i.e., 
$$
\bL^{*}(I)=I.
$$ 
\end{Theorem}

Let us return to the operational distribution $\{\bX(x)|\ x\in\R\}$ 
of the apparatus $\bA(\bx)$.  
Let $\bA(\bb)$ be an apparatus measuring a discrete observable $B$ 
with eigenvalues $b_{m}$ and 
let $\rh$ be an arbitary density operator.
By the property of joint probability, we have
$$
0\le\Pr\{\bx=x,\bb=b_{m}\|\rh\}\le 1.
$$
From \eq{ob-joint} we have
$$
0\le\Tr[E^{B}(b_{m})\bX(x)\rh]\le 1.
$$
Since $B$ and $\rh$ are arbitrary, 
the operation $\bX(x)$ is a positive superoperator.
Taking $B=I$ and $b_{m}=1$, we have
$$
0\le\Tr[\bX(x)\rh]\le 1.
$$
It follows from Thorem \ref{th:0306a} that
the operation $\bX(x)$ is a positive contractive superoperator.
By the unicity of total probability, we have
$$
\sum_{x\in\R}\Pr\{\bx=x\|\rh\}=1.
$$
Hence, we have
$$
\Tr[\sum_{x\in\R}\bX(x)\rh]=1.
$$
for all density operator $\rh$.  
Let $\bX(x)^{*}$ be the dual of the operation $\bX(x)$.
It follows that 
\beql{0307a}
\sum_{x\in\R}\bX(x)^{*}I=I.
\eeq
and that 
\beql{0307b}
\Tr[\sum_{x\in\R}\bX(x)\rh]=\Tr[\rh]
\eeq
for all $\rh\in\tc(cH)$.  
For any $x\in\R$, define the operator $X(x)$ by
\beql{effect1}
X(x)=\bX(x)^{*}I.
\eeq
We call $X(x)$ the {\em effect} of $\bA(\bx)$ for the outcome
$\bx=x$.
The family $\{X(x)|\ x\in\R\}$ of the effects of $\bA(\bx)$ is
called the {\em effect distribution} of the apparatus $\bA(\bx)$.
From  \eq{0117a} the output distribution of the apparatus $\bA(\bx)$
is determined by the effect as
\beql{effect2}
\Pr\{\bx=x\|\rh\}=\Tr[X(x)\rh].
\eeq
By the positivity of probability, we have $\Tr[X(x)\rh]\ge 0$.
Since the density operator $\rh$ is arbitrary, $X(x)$ is a positive
operator.
From \eq{0307a}, we have
\beql{effect4}
\sum_{x\in\R}X(x)=I.
\eeq
In this case, $X(x)=0$ except for countable number of $x$s.
From \eq{effect2}, $\bA(\bx)$ measures an observable $A$ 
if and only if
\beql{effect5}
X(x)=E^{A}(x).
\eeq
Thus, $\bA(\bx)$ measures an observable if and only if
the effect distribution coincides with its spectral projections.
Otherwise, the apparatus $\bA(\bx)$ is interpreted to carry out a
more general measurement such as an approximate measurement 
of an observable.

We define the positive superoperator $\bT$ by
\beql{6.3}
\bT\rh=\sum_{x\in\R}\bX(x)\rh,
\eeq
where the sum is a countable sum since $\bX(x)=0$ except for 
countable number of $x$s.
This $\bT$ is called the {\em nonselective operation} of the 
apparatus  $\bA(\bx)$.
From  \eq{effect4} we have $\bT^{*}I=I$ and hence 
$\bT$ is a trace preserving positive superoperator.

A family $\{\bW(x)|\ x\in\R\}$ of positive superoperators is called
a {\em superoperator distribution} iff
$$
\sum_{x\in\R}\bW(x)^{*}I=I.
$$
A family $\{F(x)|\ x\in\R\}$ of positive operators is called
a {\em operator distribution} iff
$$
\sum_{x\in\R}F(x)=I.
$$
The family $\{E^{A}(x)|\ x\in\}$ of spectral projections of a
discrete self-adjoint operator $A$ is an operator distribution.
The family $\{W(x)|\ x\in\R\}$ of positive operators defined by
$$
W(x)=\bW(x)^{*}I
$$
is an operator distribution and is called the 
{\em operator distribution} of $\{\bW(x)|\ x\in\R\}$.
The superoperator $\bT$ definied by
$$
\bS=\sum_{x\in\R}\bW(x)
$$
is a positive trace preserving superoperator and 
is called the {\em total superoperator} of $\{\bW(x)|\ x\in\R\}$.

We have shown under the mixing law of the joint probability that
the operational distribution $\{\bX(x)|\ x\in\R\}$
of an apparatus $\bA(\bx)$ is a superoperator distribution, 
the effect distribution of $\bA(x)$ is the operator distribution 
$\{\bX(x)|\ \rh\}$, and the nonselective superoperator of $\bA(\bx)$
is the total superoperator of $\{\bX(x)|\ x\in\R\}$.
Conversely, if for given apparatuses $\bA(\bx)$ and $\bA(\by)$ 
there are superoperator distributions $\{\bX(x)|\ x\in\R\}$
and $\{\bY(y)|\ y\in\R\}$ satisfy \eq{h} respectively, 
then the joint probability distribution 
of the outcomes of the successive measurements carried out
by $\bA(\bx)$ and $\bA(\by)$ satisfies
$$
\Pr\{\bx=x,\by=y\|\rh\}=\Tr[\bY(y)\bX(x)\rh],
$$
and hence the mixing law of the joint probability holds.

From the arguments so far,
we conclude that the mixiing law of the joint porbability is
equivalent with the following requirement: {\em For any measuring
apparatus $\bA(\bx)$, there is a superoperator distribution 
$\{\bX(x)|\ x\in\R\}$ such that the statistical property of 
$\bA(\bx)$ is represented as follows.}
\beqa
\mb{output distribution: }& &\Pr\{\bx=x\|\rh\}=\Tr[\bX(x)\rh]
\label{eq:0117a'}\\
\mb{output state: }& &\rh_{\{\bx=x\}}
=\frac{\bX(x)\rh}{\Tr[\bX(x)\rh]}\label{eq:0117b'}
\eeqa
In \eq{0117b'} the outcome $\bx=x$ is supposed to have 
positive probability; henceforce, the analogous assumption will be required 
implicitly in the similar expressions on the output state.

It follows that the problem as to what statistical property
is possible is reduced to the problem as
to what superoperator distributions are the operational distributions 
of apparatuses.

\section{Davies-Lewis postulate}

For the case of the discrete output variables,
the notion of superoperator distributions is equivalent
to the notion of normalized positive superoperator valued measures
introduced by Davies and Lewis \cite{DL70}.
A {\em positive superoperator valued (PSV) measure}
is a mapping $\cE$ which maps every Borel 
set $\De$ to a positive superoperator $\cE(\De)$
such that if $\De_{1},\De_{2},\ldots$ is a countable Borel
partition of $\De$, then we have
$$
\cE(\De)\rh=\sum_{n}\cE(\De_{n})\rh
$$
for any $\rh\in\tc(\cH)$, where the sum is convergent in the trace norm.
The PSV measure $\cE$ is said to be {\em normalized} if it satisfies
the further condition  
$$
\Tr[\cE(\R)\rh]=\Tr[\rh]
$$ 
for any $\rh\in\tc(\cH)$.
The equivalence is given below 
analogous to the case of discrete probability measures.
If $\cE$ is a normalized PSV measure, then the corresponding
superoperator distribution $\{\bX(x)|\ x\in\R\}$
is given by 
\beql{UFPM}
\bX(x)=\cE(\{x\}),
\eeq 
where $\{x\}$ is the singleton set
containing the point $x$.  
Conversely, if $\{\bX(x)|\ x\in\R\}$ is a superoperator distribution,
then the corresponding normalized PSV measure is given by
\beql{NPSVM}
\cE(\De)=\sum_{x\in\De}\bX(x).
\eeq 

For the apparatus $\bA(\bx)$, the probability $\Pr\{\bx\in\De\|\rh\}$
of obtaining the outcome in the Borel set $\De$ is given by
\beql{OD}
\Pr\{\bx\in\De\|\rh\}=\sum_{x\in\De}\Pr\{\bx=x\|\rh\}
\eeq
and the output state of the ensemble of the samples with the 
outcome in the Borel set $\De$ is given by  
\beql{OS}
\rh_{\{\bx\in\De\}}
=
\frac{\sum_{x\in\De}\Pr\{\bx=x\|\rh\}\rh_{\{\bx=x\}}}
{\Pr\{\bx\in\De\|\rh\}}.
\eeq

Davies and Lewis \cite{DL70}
proposed the following description of measurement statistics:
\vskip\topsep

{\bf Davies-Lewis postulate:}
{\em For any measuring apparatus $\bA(\bx)$, there is a normalized
PSV measure $\cE$ satisfying the following relations
for any density operator $\rh$ and Borel set $\De$:}
\vskip\topsep

(DL1) $\Pr\{\bx\in\De\|\rh\}=\Tr[\cE(\De)\rh]$
\vskip\topsep

(DL2) ${\displaystyle\rh_{\{\bx\in\De\}}=\frac{\cE(\De)\rh}
                                          {\Tr[\cE(\De)\rh]}}$
\vskip\topsep

Although the Davies-Lewis description of measurement is quite general, 
it is not clear by itself whether it is general enough to 
exhaust all the possible measurements.
Our arguments are about to complete proving the following theorem that 
shows indeed it is the case.

\begin{Theorem}\label{th:DL-ML}
The Davies-Lewis postulate is equivalent to the mixing law of the
joint probability.
\end{Theorem}

In fact, 
under the Davies-Lewis postulate, we have the normalized PSV measures
$\cE_{\bx}$ and $\cE_{\by}$ for any apparatuses $\bA(\bx)$ and $\bA(\by)$.  
By substituting (DL1) and (DL2) in \eq{c}, the joint probability is given by 
$$
\Pr\{\bx=x,\by=y\|\rh\}=\Tr[\cE_{\by}(\{y\})\cE_{\bx}(\{x\})\rh].
$$
From the linearity of $\cE_{\bx}(\{x\})$ and 
$\cE_{\by}(\{x\})$, the mixing law follows.
Conversely, under the mixing law, we have shown that there is a 
superoperator distribution $\{\bX(x)|\ x\in\R\}$  satisfying \eq{0117a}
and \eq{0117b}. 
Now, it is easy to check 
that relations \eq{UFPM}--\eq{OS} leads to the Davies-Lewis 
description (DL1)--(DL2) and the proof is completed.

\section{Measurements of discrete observables}

For a given discrete self-adjoint operator $A$, 
a superoperator distribution $\{\bX(x)|\ x\in\R\}$
is called {\em $A$-compatible} iff
$
\bX(x)^{*}I=E^{A}(x)
$
for all $x\in\R$.
The operational distribution of an apparatus measuring the
observable $A$ is an $A$-compatible superoperator distribution .

We have the following theorem \cite{97OQ}; a simplified proof
will be given in Appendix \ref{se:B}.

\begin{Theorem}\label{TH:DECOMP}
Let $A$ be a discrete self-adjoint operator.
Let $\{\bX(x)|\ x\in\R\}$ be an $A$-compatible superoperator distribution
and $\bT$ its total superoperator.
For any real number $x$ and trace class operator $\rh$, we have
\beqa\label{eq:0117c}
\bX(x)\rh
&=&\bT[E^{A}(x)\rh]=\bT[\rh E^{A}(x)]\nn\\
&=&\bT[E^{A}(x)\rh E^{A}(x)].
\eeqa
For any real number $x$ and bounded operator $B$, we have
\beqa\label{eq:0117d}
\bX(x)^{*}B
&=&E^{A}(x)\bT^{*}(B)=\bT^{*}(B)E^{A}(x)\nn\\
&=&E^{A}(x)\bT^{*}(B)E^{A}(x).
\eeqa
\end{Theorem}

From the above theorem, the operational distribution of an
apparatus measuring an observable $A$ is determined uniquely
by the nonselective operation.
It follows from \eq{0117d} that the range of $\bT^{*}$ consists
of operators commuting with $A$.
Let us define the {\em commutant} of $A$, denoted by $\{A\}'$,
as the set of all bounded operators commuting with $A$.
A trace preserving positive superoperator $\bL$ on $\tc(\cH)$
is called {\em $A$-compatible} iff the range of its duel $\bL^{*}$
is included in the commutant $\{A\}'$ of $A$.

For any trace preserving positive superoperator $\bL$, let
$$
\bL'\rh=\sum_{x\in\R}\bL[E^{A}(x)\rh E^{A}(x)].
$$
Then $\bL'$ is an $A$-compatible positive superoperator.
Obviously, $\bL$ itself is $A$-compatible if and only if $\bL'=\bL$.

From \eq{0117d}, the total superoperator of an $A$-compatible 
superoperator distribution is an $A$-compatible positive superoperator.
Conversely, for any $A$-compatible positive superoperator $\bT$,
let $\bX(x)\rh=\bT[E^{A}(x)\rh]$ for all $\rh\in\tc(\cH)$.
Then $\{\bX(x)|\ x\in\R\}$ is an $A$-compatible superoperator 
distribution and $\bT$ is its total superoperator.
From the above argument, we have obtained the following theorem.

\begin{Theorem}\label{th:A-COMP}
Let $A$ be a discrete self-adjoint operator on $\cH$.
The relation
\beql{compatible}
\bX(x)\rh=\bT[E^{A}(x)\rh]
\eeq
for all real number $x$ and trace class operator $\rh$
sets up a one-to-one correspondence between 
the $A$-compatible superoperator distribution $\{\bX(x)|\ x\in\R\}$
and the $A$-compatible positive superoperators $\bT$.
\end{Theorem}

From the above theorem, we conclude the following:
{\em For any apparatus $\bA(\bx)$ measuring a discrete observable $A$, 
there is an $A$-compatible positive superoperator $\bT$ 
such that the statistical property of $\bA(\bx)$ is represented as follows.}
\beqa
\mb{output distribution: }& &\Pr\{\bx=x\|\rh\}=\Tr[E^{A}\rh]\label{eq:0117e}\\
\mb{output state: }& &\rh_{\{\bx=x\}}
=\frac{\bT[E^{A}(x)\rh]}{\Tr[E^{A}(x)\rh]}\label{eq:0117f}
\eeqa

It follows that the problem of determining all the possible quantum
state reductions $\rh\to\rh_{\{\bx=x\}}$ arising in the apparatus
measuring $A$ is reduced to the following problems: (i) Does every
$A$-compatible positive superoperator have the corresponding measuring
apparatus? (ii) If not, what condition does ensure the existence
of the corresponding measuring apparatus?

\section{Measurements of nondegenerate discrete observables}

In this section, we confine our attention to the observables 
with nondegenerate eigenvalues.
In this case, the projection $E^{A}(a_{n})$ is of rank 1
and is the density operator representing the eigenstate,
so that we have
$$
E^{A}(a_{n})\rh E^{A}(a_{n})=\Tr[E^{A}(a_{n})\rh]E^{A}(a_{n}).
$$
Let $\bT$ be an $A$-compatible positive superoperator.   
From \eq{0117c}, we have
\beql{0118a}
\bT[E^{A}(a_{n})\rh]=\Tr[E^{A}(a_{n})\rh]\bT[E^{A}(a_{n})].
\eeq
We define a sequence $\{\bold{\vr}_{n}\}$ of density operators by
\beql{301b}
\bold{\vr}_{n}=\bT[E^{A}(a_{n})].
\eeq
Then, we have
\beql{0120a}
\bT(\rh)=\sum_{n}\Tr[E^{A}(a_{n})\rh]\bold{\vr}_{n}.
\eeq
Conversely, for any sequence $\{\bold{\vr}_{n}\}$ of density operators,
we define the positive superoperator $\bT$ on $\tc(\cH)$ by \eq{0120a}.
Then, $\bT$ is an $A$-compatible positive superoperator satisfying \eq{301b}.
Thus, we have proved the following theorem.

\begin{Theorem}\label{th:ND}
Let $A$ be a nondegenerate discrete self-adjoint operator.
The relation
$$
\bT(\rh)=\sum_{n}\Tr[E^{A}(a_{n})\rh]\bold{\vr}_{n},
$$
where $\rh\in\tc(\cH)$, sets up a one-to-one correspondence
between  the families $\{\bvr_{x}|\ x\in\R\}$ of density operators
and the $A$-compatible positive superoperators $\bT$ on $\tc(\cH)$.
\end{Theorem}

Let $\rh_{\{\bx=x\}}$ be the output state of an apparatus measuring
$A$ for the input state $\rh$.
Then, there is an $A$-compatible positive superoperator $\bT$ satisfying 
\eq{0117f} and there is a sequence $\{\bvr_{n}\}$ of
density operators satisfying \eq{301b}, so that we have
$$
\rh_{\{\bx=a_{n}\}}=\frac{\bT[E^{A}(a_{n})\rh]}{\Tr[E^{A}(a_{n})\rh]}
=\bT[E^{A}(a_{n})]=\bvr_{n}.
$$
It follows that the output
state for the output $\bx=a_{n}$ is given by
\beq
\rh_{\{\bx=a_{n}\}}=\bvr_{n}.
\eeq

From the above argument we conclude the following:
{\em For any apparatus $\bA(\bx)$ measuring a nondegenerate discrete 
observable 
$A=\sum_{n}a_{n}\ket{\ph_{n}}\bra{\ph_{n}}$, there is 
a sequence $\{\bold{\vr_{n}}\}$ of density operators such that
the statisitcal property of $\bA(\bx)$ is represented 
as follows.}
\beqa
\mb{output distribution: }& &\Pr\{\bx=a_{n}\|\rh\}
=\bracket{\ph_{n}|\rh|\ph_{n}}\label{eq:nondeg1}\\
\mb{output state: }& &\rh_{\{\bx=a_{n}\}}=\bold{\vr}_{n}        
\label{eq:nondeg2}
\eeqa

It follows that the problem of determining all
the possible quantum state reductions arising in the measurement
of a nondegenerate discrete observable $A$ is reduced to the problem as
to what sequence $\{\bold{\vr}_{n}\}$ of states can be obtained from
the measurement of $A$.
In order to obtain the answer to this question,
in the next section we shall consider indirect measurement models
and ask what sequences can be obtained from those models.

It should be noted here that the apparatus satisfies the projection 
postulate if and only if
we have 
$$
\bold{\vr_{n}}=E^{A}(a_{n})
$$ 
for all $n$.
Von Neumann \cite[pp.~439--442]{vN55}
showed that this case can be obtained from an indirect
measurement model.

\section{Indirect measurement models}\label{se:measuring}

In general, if a measurement on the object in the input state $\rh$
by the apparatus $\bA(\bx)$ is immediately followed by
a measurement of the observable $B$ by the apparatus $\bA(\bb)$, 
the joint probability distribution of their output variables
is given by \eq{ob-joint}.
Now, consider the marginal probability 
\beql{f'}
\Pr\{\bx\in\R,\bb=b_{m}\|\rh\}=\sum_{x\in\R}\Pr\{\bx=x,\bb=b_{m}\|\rh\}.
\eeq
Then, this represents the probability of obtaining the outcome $\bb=b_{m}$
after interacting the apparatus $\bA(\bx)$ with the object 
without reading out the outcome of the $\bx$ measurement.
Such a process of is called a {\em nonselective measurement}.
Let $\bT$ be the nonselective operation of the apparatus $\bA(\bx)$.
Then, by \eq{ob-joint}, we have
\beq
\Pr\{\bx\in\R,\bb=b_{m}\|\rh\}
=
\Tr[E^{B}(b_{m})\bT\rh].
\eeq
Thus, the nonselective measurement transforms the input state $\rh$
to the output state $\rh_{\{\bx\in\R\}}=\bT\rh$.

Let us call any interaction between the object and the apparatus
caused by a measurement as the {\em measuring interaction}.
Then, the superoperator $\bT$ is determined by the measuring interaction.
In what follows we shall examine the properties of the measuring
interaction.

Since the nonselective measurement transforms the input
state $\rh$ to the output state $\bT\rh$, 
there should be an interaction
during finite time interval when the object changes from
$\rh$ to $\bT\rh$.
Moreover, the object should be free from the apparatus
before and after the interaction.
Thus, we suppose that the measuring interaction turns on from
the time $t$ just before the measurement to the time
$t+\De t$ just after the measurement where $\De t>0$,
and that the object is free from the apparatus before the 
the time $t$ and after the time $t+\De t$.
It follows that if the second measurement on the same object
follows immediately after the above measurement, the time just
before the second measurement coincides with the time 
$t+\De t$ just after the first measurement.
In this way, the temporal boundary of the measuring interaction 
is determined as a fixed domain from time $t$ to $t+\De t$.

Next, in order to determine the spatial boundary of the measuring
interaction, we consider the smallest subsystem of the measuring
apparatus such that the composite system of the object and
the subsystem is isolated from the time $t$ to the time $t+\De t$.
We call the above subsystem as the {\em prove}.

The effect of the measuring interaction is given by the change
of an observable $M$, called the {\em probe observable},
from $t$ to $t+\De t$.
From the minimality of the probe, it is natural to assume that 
the interaction Hamiltonian excludes any macroscopic part of 
the measuring apparatus such as the macroscopic pointer position.
It follows that the measuring interaction is a quantum mechanical
interaction and the state change can be described by the unitary
time evolution of the composite system of the object and
the probe.

On the other hand, in order to transduce the microscopic change in 
the probe observable $M$ to the macroscopic change such as the change
of the position of the pointer, we need an amplification process
in the apparatus after $t+\De t$.
This transduction from a microscopic observable to a macroscopic
observable corresponds to the direct measurement of the probe 
observable $M$ at the time $t+\De t$.
The problem of describing this process as a dynamical process
belongs to the so-called measurement problem.
Within quantum mechanics, the Born statistical formula gives
the the probability distribution of the outcome of the
$M$ measurement.  Let $t+\De t+\ta$ be the time just after
this amplification process where $\ta>0$.
This time is called {\em the time of read-out}.

According to the above description,
the process from the time just before the measurement to the 
time of read-out is divided into the measuring interaction 
and the amplification.
It should be noted that just after the measuring interaction, 
the object is free from the apparatus so that it is possible to start the 
interaction with the second apparatus.
It follows that in the successive measurement experiment
the time just before the second measurement is 
considered to be the time just after the measuring interaction 
rather than the time of read-out.
The above description of measuring process is called 
an {\em indirect measurement description}.

Let $\cH$ be the state space of the object $\bS$, 
and $\cK$ the state space of the probe $\bP$.
The state of the object at the time $t$ of measurement
is the input state $\rh$.
The probe $\bP$ is supposed to be prepared in the fixed state
$\si$ at the time of measurement.
Thus, the state of the composite system at the time $t$ is
$$
\bold{\rh}_{\bS+\bP}(t)=\rh\otimes\si.
$$
If the time evolution of the composite system $\bS+\bP$ from 
$t$ to $t+\De t$ is represented by the unitary operator $U$,
the composite system is in the state 
\beql{828d}
\bold{\rh}_{\bS+\bP}(t+\De t)=U(\rh\otimes\si)U^{\da}
\eeq
at $t+\De t$.
Suppose that the $\bA(\bx)$ measurement in $\rh$ is followed 
immediately by a measurement of an observable $B$ carried out 
by $\bA(\bb)$.
Then, the observable $B$ is measured at the time $t+\De t$ and
the outcome is recorded by $\bb$.
On the other hand, the probe observable $M$ is also measured actually
at the time $t+\De t$ and the outcome is recorded by $\bx$. 
Since the two measurements are carried out locally, it follows from
the local measurement theorem  \cite{97QQ,98QS} that 
the joint probability distribution of the outcomes of the above 
two  measurements satisfies
\beqa
\lefteqn{\Pr\{\bx=x,\bb=b_{m}\|\rh\}}\quad\nn\\
&=&
\Tr[(E^{B}(b_{m})\otimes E^{M}(x))U(\rh\otimes\si)U^{\da}]\nn\\
&=&
\Tr[E^{B}(b_{m})
\Tr_{\cK}[(I\otimes E^{M}(x))U(\rh\otimes\si)U^{\da}]],
\eeqa
where $\Tr_{\cK}$ is the partial trace over the Hilbert space
$\cK$.  Thus, from \eq{ob-joint} we have
\beql{meas-pro}
\bX(x)\rh=\Tr_{\cK}[(I\otimes E^{M}(x))U(\rh\otimes\si)U^{\da}].
\eeq
Hence, the statistical property of the apparatus
$\bA(\bx)$ is given as follows.
\beqa
\lefteqn{\mb{output distribution: }}\qquad\nn\\
& &
\Pr\{\bx=x\|\rh\}=\Tr[(I\otimes E^{M}(x))U(\rh\otimes\si)U^{\da}]
\label{eq:828e}
\\
\lefteqn{\mb{output state: }}\qquad\nn\\
& &\rh_{\{\bx=x\}}=
\frac{\Tr_{\cK}[(I\otimes E^{A}(x))U(\rh\otimes\si)U^{\da}]}
           {\Tr[(I\otimes E^{A}(x))U(\rh\otimes\si)U^{\da}]}
\eeqa
From \eq{6.3} and \eq{meas-pro}, 
the nonselective operation of $\bA(\bx)$ is given by
\beql{829d}
\bT\rh=\Tr_{\cK}[U(\rh\otimes\si)U^{\da}].
\eeq
From \eq{effect2} and \eq{828e}, the effect distribution of $\bA(\bx)$ is
given by
\beq
X(x)=\Tr_{\cK}[U^{\da}(I\otimes E^{M}(x))U(I\otimes\si)].
\eeq

In general, a four tuple $(\cK,\si,U,M)$ is called an {\em
indirect measurement model} iff it consists of 
a separable Hilbert space $\cK$$B!$(B
a density operator $\si$ on $\cK$,
a unitary operator $U$ on $\cH\otimes\cK$,
and a self-adjoint operator $M$ on $\cK$.
So far we have not posed any sufficient condition for the existence
of an apparatus except that every observable has at least one
apparatus to measure it. 
Here, we pose the following hypothesis.
\vskip\topsep

{\bf Unitary realizability hypothesis: }
{\em For any indirect measurement model $(\cK,\si,U,M)$, there is an 
apparatus $\bA(\bx)$ with the following statistical property:}
\beqas
\lefteqn{\mb{output distribution: }}\qquad\nn\\
& &
\Pr\{\bx=x\|\rh\}=\Tr[(I\otimes E^{M}(x))U(\rh\otimes\si)U^{\da}]\\
\lefteqn{\mb{output state: }}\qquad\nn\\
& &\rh_{\{\bx=x\}}=
\frac{\Tr_{\cK}[(I\otimes E^{A}(x))U(\rh\otimes\si)U^{\da}]}
           {\Tr[(I\otimes E^{A}(x))U(\rh\otimes\si)U^{\da}]}
\eeqas

A superoperator distribution $\{\bX(x)|\ x\in\R\}$ is said to 
be {\em realized} by an indirect measurement model $(\cK,\si,U,M)$ iff 
\eq{meas-pro} holds for any $\rh\in\tc(\cH)$,  and in this case
it is called {\em unitarily realizable}.
Under the unitary realizability hypothesis, unitarily realizable
superoperator distributions are operational distributions of
some apparatuses.
In the next section, we shall give an intrinsic characterization of 
the unitarily realizable superoperator distributions.

\section{Complete positivity}

Let $\cD=\tc(\cH)$ or $\cD=\cL(\cH)$.
A linear transformation $\bL$ on $\cD$
is called {\em completely positive (CP)} iff for any finite
sequences of bounded operators $A_{1},\ldots,A_{n}\in\cD$
and vectors $\xi_{1},\ldots,\xi_{n}\in\cH$ we have
$$
\sum_{ij}\bracket{\xi_{i}|\bL(A_{i}^{\da}A_{j})|\xi_{j}}\ge0.
$$
The above condition is equivalent to that $\bL\otimes I$
maps positive operators in the algebraic tensor product
$\cD\otimes\cL(\cK)$ to positive operators in 
$\cD\otimes\cL(\cK)$ for any Hilbert space $\cK$.
Obviously, every CP superoperators are positive.
A superoperator is CP if and only if its dual superoperator is CP.
A superoperator distribution $\{\bX(x)|\ x\in\R\}$ is called
{\em completely positive} iff every $\bX(x)$ is CP.
It can be seen easily from \eq{meas-pro} that unitarily
realizable superoperator distributions are CP.
Conversely, the following theorem, proved in \cite{83CR,84QC}
for an even more general formulation, asserts that every 
CP superoperator distribution is unitarily realizable.

\begin{Theorem}\label{th:rep}
For any CP superoperator distribution $\{\bX(x)|\ x\in\R\}$,
there is a separable Hilbert space $\cK$,
a unit vector $\Ph$ in $\cK$, 
a unitary operator $U$ on $\cH\otimes\cK$,
and a discrete self-adjoint operator $M$ on $\cK$
satisfying the relation
$$
\bX(x)\rh=\Tr_{\cK}[(I\otimes E^{M}(x))U(\rh\otimes\si)U^{\da}].
$$
for all $\rh\in\tc(cH)$.
\end{Theorem}

For any trace preserving CP superoperator $\bT$, 
we have a CP superoperator distribution 
$\{\bX(x)|\ x\in\R\}$ such that $\bX(0)=\bT$ and
that $\bX(x)=0$ for all $x\ne 0$.  Applying the above theorem
to this family, we obtain the following representation theorem of
trace preserving CP superoperators, 
which was proved independently by Kraus \cite{Kra83} and the
present author \cite{83CR}.

\begin{Theorem}\label{th:rep-CP}
For any trace preserving CP superoperator $\bT$,
there is a separable Hilbert space $\cK$,
a unit vector $\Ph$ in $\cK$, 
a unitary operator $U$ on $\cH\otimes\cK$,
such that $\bT$ satisfies the relation
$$
\bT\rh=\Tr_{\cK}[U(\rh\otimes\ket{\Ph}\bra{\Ph})U^{\da}].
$$
for all $\rh\in\tc(\cH)$.
\end{Theorem}

From Theorem \ref{TH:DECOMP}, every $A$-compatible superoperator 
distribution $\{\bX(x)|\ \in\R\}$ satisfies the relation
$\bX(x)\rh=\bT[E^{A}(x)\rh E^{A}(x)]$. 
Thus, if $\{\bX(x)|\ x\in\R\}$ is CP, 
then the total map $\bT=\sum_{x\in\R}\bX(x)$ is CP, since
the sum of CP superoperators is CP.  
Conversely, if $\bT$ is an $A$-compatible CP superoperator, 
then the corresponding $A$-compatible superoperator distribution
$\{\bX(x)|\ x\in\R\}$ is CP, since the
superoperator $\rh\mapsto E^{A}(x)\rh E^{A}(x)$ is CP and the composition
of any CP superoperators is CP.
Thus we have the following:

\begin{Theorem}\label{th:A-COMP-CP}
Let $A$ be a nondegenerate discrete self-adjoint operator.
Then, an $A$-compatible superoprator distribution is CP 
if and only if its total superoperator is CP.
\end{Theorem}

From the above theorem, we conclude the following \cite{84QC}:
{\em The statistical equivalence classes of apparatuses $\bA(\bx)$ 
measuring a discrete observable $A$ with indirect measurement models are
in one-to-one correspondence with the $A$-compatible CP superoperators,
where the statistical property is represented by \eq{0117e} and \eq{0117f}.}

Now, let $A$ be a nondegenerate discrete observable and
let $\bT$ be an $A$-compatible positive superoperator.
Then, $\bT$ is of the form \eq{0120a}.  Let 
$\si_{1},\ldots,\si_{n}\in\tc(\cH)$
and $\xi_{1},\ldots,\xi_{n}\in\cH$. 
Then, we have
\beqas
\sum_{ij}\bracket{\xi_{i}|\bT(\si_{i}^{\da}\si_{j})|\xi_{j}}
&=&
\sum_{n}\sum_{ij}\Tr[E^{A}(a_{n})\si_{i}^{\da}\si_{j}]
	\bracket{\xi_{i}|\bvr_{n}|\xi_{j}}\\
&\ge& 0,
\eeqas
where the last inequality follows from the fact that the trace of
the product of two positive definite matrices 
$(\Tr[E^{A}(a_{n})\si_{i}^{\da}\si_{j}])_{ij}$ and 
$(\bracket{\xi_{i}|\bvr_{n}|\xi_{j}})_{ij}$ is nonnegative. 
It follows that $\bT$ is a CP superoperator.
Thus, every $A$-compatible superoperator is CP.
Since every $A$-compatible superoperator distribution is obtained 
from an $A$-compatible superoperator by Theorem \ref{th:A-COMP},
it follows from Theorem \ref{th:A-COMP-CP} that
every $A$-compatible superoperator distribution is CP.
We have therefore obtained the following statements.

\begin{Theorem}
Let $A$ be a nondegenerate discrete self-adjoint operator.
Every $A$-compatible positive superoperator is completely 
positive.  Every $A$-compatible superoperator distribution
is completely positive.
\end{Theorem}

From the above theorem and Theorem \ref{th:rep} we conclude:
{\em Every apparatus measuring $A$ is statistically equivalent to the one 
having an indirect measurement model.}

Every sequence $\{\bvr_{n}\}$ of density operators defines an
$A$-compatible positive superoperator by Theorem \ref{th:ND},
and it is automatically completely positive so that it is
realized by an idirect measurement model.
Thus, we have reached the answer to the question what sequence of states
can be obtained from an apparatus measuring $A$ that every
sequence can.  Thus, we conclude the following:
{\em   The statistical equivalence classes of apparatuses $\bA(\bx)$ 
measuring a nondegenerate discrete observable $A$ are in one-to-one
correspondence with the sequences $\{\bvr_{n}\}$ of density operators,
where the statistical property is represented by \eq{nondeg1} and 
\eq{nondeg2}.}

\sloppy
Given any sequence $\{\bvr_{n}\}$, an indirect measurement model
with the quantum state reduction 
$$
\rh\mapsto\rh_{\{\bx=a_{n}\}}=\bvr_{n}
$$ 
is constructed explicitly as follows.
Let $\{\ph_{n}\}$  be an orthonormal basis of $\cH$ consisting of the 
eigenvectors of $A$.
Let $\cK=\cH\otimes\cH$.
Let
$$
\bvr_{n}=\sum_{j}\la_{nj}\ket{\et_{nj}}\bra{\et_{nj}}
$$
be the spectral decomposition of $\bvr_{n}$.
Then, there exists a unitary operator $U$ on $\cH\otimes\cK$ 
satisfying
$$
U\ket{\ph_{n}\otimes\ph_{0}\otimes\ph_{0}}
=
\sum_{j}\sqrt{\la_{nj}}\ket{\et_{nj}\otimes\ph_{j}\otimes\ph_{n}}.
$$
Now, we define the density operator $\si$ on $\cK$ by
$\si=\ket{\ph_{0}\otimes\ph_{0}}\bra{\ph_{0}\otimes\ph_{0}}$
and define a self-adjoint operator $M$ on $\cK$ by 
$M=I\otimes A$.
Then, we have the indirect measurement model $(\cK,\si,U,M)$
such that the statistical property of its apparatus
satisfies \eq{nondeg1} and \eq{nondeg2}.

\section{Conclusions}

Let $\bA(\bx)$ be an apparatus with the discrete output variable
$\bx$.  
Then, depending on the input state $\rh$ and the outcome $x$,
the apparatus $\bA(\bx)$ determines the output probability $\Pr\{\bx=x\|\rh\}$ 
and the output state $\rh_{\{\bx=x\}}$.
The transformation from the input state $\rh$ to the output 
distribution $\Pr\{\bx=x\|\rh\}$ is called 
the objective state reduction and 
the one from the input state $\rh$ to the  output 
states $\rh_{\{\bx=x\}}$ is called 
the quantum state reduction.
The pair of the objective state reduction and the quantum state reduction
is called the statistical property of the apparatus $\bA(\bx)$.
Two apparatuses with the same statistical property is said to be
statistically equivalent.
In order to obtain a mathematical description of quantum state 
reductions for the most general class of measurements 
we have considered two requirements:
one is necessary and the other is sufficient.

The necessary one is the mixing law of the joint probability.
Suppose that a measurement carried out by an apparatus 
$\bA(\bx)$ in the input state $\rh$ is followed immediately 
by another measurement carried out by another apparatus $\bA(\by)$.
The joint probability distribution of the outcomes 
$\bx$ and $\by$ is determined by their statistical properties
as follows.
$$
\Pr\{\bx=x,\by=y\|\rh\}
=\Pr\{\by=y\|\rh_{\{\bx=x\}}\}\Pr\{\bx=x\|\rh\}.
$$
This joint probability distribution is considered to respect
the mixture of input states and the mixing law of the joint probability
requires that this is the case 
for any apparatuses $\bA(\bx)$ and $\bA(\by)$.
Under this hypothesis, any apparatus $\bA(\bx)$ has a
superoperator distribution $\{\bX(x)|\ x\in\R\}$, 
called the operational distribution of $\bA(\bx)$,
satisfying
\beql{0308a}
\bX(x)\rh=\Pr\{\bx=a\|\rh\}\rh_{\{\bx=x\}}.
\eeq

The sufficient condition is the unitary realizability condition.
The apparatus $\bA(\bx)$ is said to have an indirect measurement model 
$(\cK,\si,U,M)$ iff the statistical property of $\bA(\bx)$ is given
as follows.
\beqas
\lefteqn{\mb{output distribution: }}\qquad\\
& &\Pr\{\bx=x\|\rh\}=
\Tr[(I\otimes E^{M}(x))U(\rh\otimes\si)U^{\da}]\\
\lefteqn{\mb{output state: }}\qquad\\
& &\rh_{\{\bx=x\}}=
\frac{\Tr_{\cK}[(I\otimes E^{M}(x))U(\rh\otimes\si)U^{\da}]}
           {\Tr[(I\otimes E^{M}(x))U(\rh\otimes\si)U^{\da}]}
\eeqas
In general, an apparatus has an indirect measurement model if and only
if its operational distribution is completely positive.
The unitary realizability hypothesis states that every indirect
measurement model defines an apparatus with the above statistical
property.
It follows that the statistical equivalence classes of 
apparatuses with indirect measurement models are in one-to-one 
correspondence with the CP superoperator distributions 
$\{\bx(x)|\ x\in\R\}$, under the relation \eq{0308a}.

Let $A$ be a discrete observable.
A trace preserving positive superoperator $\bL$
is called $A$-compatible iff the range of its duel $\bL^{*}$
is included in the commutant $\{A\}'$ of $A$.
The statistical 
property of an apparatus measuring an observable $A$ is
represented by an $A$-compatible positive superoperator $\bT$ as follows.
\beqa
\mb{output distribution: }& &\Pr\{\bx=x\|\rh\}=\Tr[E^{A}\rh]\\
\mb{output state: }& &\rh_{\{\bx=x\}}
=\frac{\bT[E^{A}(x)\rh]}{\Tr[E^{A}(x)\rh]}
\eeqa
In particular, the statistical equivalence classes of apparatuses 
with indirect measurement models measuring $A$ are in
one-to-one correspondence with the $A$-compatible completely
positive superoperators $\bT$, under the above description.

According to the above,
the class of possible quantum state reductions is included in the class 
of conditional state transformations satisfying the mixing law, i.e., 
the general superoperator distributions,
and includes the one satisfying the unitary realizability condition, 
i.e., the completely positive superoperator distributions.
Since these two classes are generally different,
there seems to be still a room for the debate in measurement 
theory on what class between them is the true class of all the possible
quantum sate reductions.

Nevertheless, for the case where $A$ is nondegenerate,
this paper shows, the above two conditions are actually equivalent.
Thus, both of them are necessary and sufficient and we reach a 
clear-cut conclusion.
In fact, if $A$ is nondegenerate, all the $A$-compatible positive
superoperators $\bT$ are completely positive 
and they are in one-to-one correspondence with the sequences 
$\{\bvr_{n}\}$ of density operators, under the relation 
$\bT(E^{A}(a_{n}))=\bvr_{n}$ where $\{a_{n}\}$ is the sequence of 
the eigenvalues of $A$.
In this case, every apparatus measuring $A$ is statistically
equivalent with the one with an indirect measurement model.
The statistical equivalence classes of the apparatuses measuring
$A$ are, therefore, 
in one-to-one correspondence with the sequences $\{\bvr_{n}\}$ 
of density operators and their statistical properties are 
represented as follows.
\beqa
\mb{output distribution: }& &\Pr\{\bx=a_{n}\|\rh\}=\Tr[E^{A}(a_{n})\rh]
\label{eq:0308b}\\
\mb{output states: }& &\rh_{\{\bx=a_{n}\}}=\bvr_{n}
\label{eq:0308c}
\eeqa

The above measurement statistics has the following two remarkable
features: (i) The output states are independent of the input state.
(ii) The family of output states can be arbitrarily chosen by the
choice of the apparatus.  The possibility of this kind of generalized
measurements was first pointed out in part by Gordon and Louisell 
\cite{GL66} relative to the measurement of an overcomplete family
of states generalizing the conventional measurement of an orthonormal 
basis.  Yuen \cite{Yue83} generalized the Gordon-Louisell description
to the following measurement described by the set of operators 
$\{\ket{\Ps_{x}}\bra{\Ph_{x}}\}$, where $\{\Ph_{x}\}$ is
an overcomplete family of vectors and $\{\Ps_{x}\}$ is a Borel
family of state vectors, as follows.
\beqas
\mb{output distribution: }& &
\Pr\{\bx\in dx\|\rh\}=\bracket{\Ph_{x}|\rh|\Ph_{x}}\,dx\\
\mb{output states: }& &\rh_{\{\bx=x\}}=\ket{\Ps_{x}}\bra{\Ps_{x}}
\eeqas
The unitary realizability of the above measurement statistics 
was assumed by Yuen \cite{Yue83} to claim the realizability of
the contractive state measurement and proved rigorously
in \cite{85CC}; see \cite{89RS} for survey.
We can see that for the nondegenerate discrete observable 
$A=\sum_{n}a_{n}\ket{\Ph_{n}}\bra{\Ph_{n}}$
and the output states $\bvr_{n}=\ket{\Ps_{n}}\bra{\Ps_{n}}$ 
the measurement statistics given in \eq{0308b} 
and \eq{0308c} corresponds to the (discrete version of)
measurement described by $\{\ket{\Ps_{n}}\bra{\Ph_{n}}\}$.  
The present paper has proved rigorously, even without assuming the 
unitary realizability, that every measurement of a nondegenerate
discrete observable is always of this form.

Along with the analogous arguments, it can be shown that the 
statistical equivalence classes of the apparatuses 
measuring a nondegenerate (but not necessarily discrete) observable
including the position or the momentum observable are in one-to-one 
correspondence with the Borel families of density
operators (modulo the spectral measure).
Since the precise mathematical formulation for that result is beyond
the scope of this paper, we shall discuss the nondiscrete case
in a separate article.

Therefore, we can conclude that as long as the statistical properties 
of measurements of nondegenerate observables are concerned, 
we can always assume that the measuring process are described by
an indirect measurement model in which the interaction between the object 
and the apparatus is described by a unitary operator.
For measurements of degenerate observables and even for
measurements of general probability operator valued measures,  
it appears to be an important question whether every 
apparatus is statistically equivalent with the one having the 
indirect measurement model that has the unitary measuring interaction. 
Since in this case there are many superoperator distributions 
(or normalized PSV measures) that are not completely positive
\cite{Cho72}, we need further physical requirements to settle 
this problem.

Following von Neumann \cite{vN55}, some authors appear
to support the hypothesis that every apparatus has 
an indirect measurement model, 
the converse of the unitary realizability hypothesis.
If this is the case, the description of measuring processes will
be simplified considerably as shown in Section \ref{se:measuring}.
In particular, we have an instant of time at which the measuring
process is divided into the measuring interaction and the amplification
process (including the so-called decoherence process) and the
output state has been prepared for the next measurement before
the amplification mode of the first measurement \cite{97OQ}.
It is also interesting whether non-conventional quantum
mechanics such as nonlinear quantum mechanics will provide a
different measurement statistics from the unitarily realizable
ones.

% APPENDIX
\appendix
\section{Linear extension of the quantum state reduction}
\label{se:A}

For any $x\in\R$ and any density operator $\rh$, the trace
class operator $\bX(x,\rh)$ is defined by \eq{f}.
In this section, we shall prove that the mapping $\bX(x):\rh\mapsto
\bX(x,\rh)$ defined on the space of density operators can be
extended uniquely to a linear transformation on the space
$\tc(\cH)$ of trace class operators on $\cH$.
By the linearity of the extension, for any trace class operator
$\si$ with decomposition \eq{T7} it is necessary for $\bX(x)\si$
to be defined by \eq{T8}.
Since the decomposition $\eq{T7}$ is not unique, in order for
the extension \eq{T8} to be well-defined we need to show that
the right hand side of \eq{T8} is uniquely determined independent
of the decomposition of $\si$.
Namely, we need to prove that {\em if $\si$ has another decomposition
\beq\label{eq:T9}
\si
=
\la'_{1}\si'_{1}-\la'_{2}\si'_{2}+i\la'_{3}\si'_{3}-i\la'_{4}\si'_{4},
\eeq
then we have}
\beqa\label{eq:T10}
\lefteqn{
\la_{1}\bX(x)\si_{1}-\la_{2}\bX(x)\si_{2}+i\la_{3}\bX(x)\si_{3}-i\la_{4}\bX(x)\si_{4}}
\nonumber\\
&=&
\la'_{1}\bX(x)\si'_{1}-\la'_{2}\bX(x)\si'_{2}
+i\la'_{3}\bX(x)\si'_{3}-i\la'_{4}\bX(x)\si'_{4}.\nonumber\\
\eeqa

The proof runs as follows \cite{Kad65}.
By equating the right hand sides of \eq{T7} and \eq{T9} and
comparing the real and imaginary parts in both sides, we have
\beqa
\la_{1}\si_{1}+\la'_{2}\si'_{2}
&=&
\la'_{1}\si'_{1}+\la_{2}\si_{2}\label{eq:T11}\\
\la_{3}\si_{3}+\la'_{4}\si'_{4}
&=&
\la'_{3}\si'_{3}+\la_{4}\si_{4}.\label{eq:T12}
\eeqa
Taking the trace of both sides of \eq{T11}, we have
\beq\label{eq:T13}
\la_{1}+\la'_{2}=\la'_{1}+\la_{2}.
\eeq
By dividing both sides of \eq{T11} by this value, we have
$$
\al\si_{1}+(1-\al)\si'_{2}
=
\be\si'_{1}+(1-\be)\si_{2},
$$
where we define $\al$ and $\be$ by
\beqas
0<\al&=&\frac{\la_{1}}{\la_{1}+\la'_{2}}<1\\
0<\be&=&\frac{\la'_{1}}{\la'_{1}+\la_{2}}<1.
\eeqas
Thus, from \eq{g} we have
$$
\al \bX(x)\si_{1}+(1-\al)\bX(x)\si'_{2}
=
\be \bX(x)\si'_{1}+(1-\be)\bX(x)\si_{2}.
$$
Multiplying both sides by the value of \eq{T13}, we have
$$
\la_{1}\bX(x)\si_{1}-\la_{2}\bX(x)\si_{2}
=
\la'_{1}\bX(x)\si'_{1}-\la'_{2}\bX(x)\si'_{2}.
$$
By the similar manipulations for \eq{T12}, we have
$$
i\la_{3}\bX(x)\si_{3}-i\la_{4}\bX(x)\si_{4}
=
i\la'_{3}\bX(x)\si'_{3}-i\la'_{4}\bX(x)\si'_{4}.
$$
Thus, we have proved equation \eq{T10}.
It is concluded, therefore, that $\bX(x)\si$ is defined uniquely
for every $\si$ by \eq{T8}.

\section{Proof of Theorem \ref{TH:DECOMP}}
\label{se:B}

Let $\{\bX(x)|\ x\in\R\}$ be an $A$ compatible family of positive
maps and $\bT$ its total map.
Let $C$ be a bounded operator such that $0\le C\le I$
and let $x\in\R$.
We define
\begin{eqnarray*}
A_{11}&=&\bX(x)^{*}C,\\              
A_{12}&=&\bX(x)^{*}(I-C),\\
A_{21}&=&\sum_{y\ne x}\bX(y)^{*}C,\\ 
A_{22}&=&\sum_{y\ne x}\bX(y)^{*}(I-C),\\
P_{1}&=&E^{A}(x),\\                  
P_{2}&=&I-E^{A}(x),\\
Q_{1}&=&\bT^{*}(C),\\                     
Q_{2}&=&I- \bT^{*}(C).
\end{eqnarray*}
Then $0\le A_{ij}\le P_{i}$, so that $[A_{ij},P_{i}]=[A_{ij},P_{j}]=0$.
It follows that $Q_{j}=A_{1j}+A_{2j}$ commutes with $P_{1}$
and $P_{2}$ as well.
Thus, 
$$
A_{ij}=P_{i}A_{ij}\le P_{i}Q_{j}.
$$
On the other hand, we have $\sum_{ij}A_{ij}=I$ and 
$\sum_{ij}P_{i}Q_{j}=I$, whence $A_{ij}=P_{i}Q_{j}$.
It follows that $\bX(x)^{*}C=E^{A}(x)\bT^{*}(C)$.
Since any bounded operator $B$ can be represented by
$B=\sum_{n=0}^{3}i^{n}\la_{n}C_{n}$ with positive
operators $0\le C_{n}\le I$
and positive reals $\la_{n}$, we have
$\bX(x)^{*}B=E^{A}(x)\bT^{*}(B)$
for any real number $x$ and bounded operator $B$.
Since $[E^{A}(x),\bT^{*}(B)]=0$, other assertions follow immediately.

% BIBLIOGRAPHY:                                                    

% END DOCUMENT:

\begin{references}\itemsep=0in

\bibitem{vN55}
J. von~Neumann,
\newblock {\it Mathematical Foundations of Quantum Mechanics}
\newblock (Princeton UP, Princeton, NJ, 1955).

\bibitem{Lud51}
G. {L\"{u}ders},
\newblock {Ann.\ Physik (6)} {\bf 8} (1951), {322}.

\bibitem{DL70}
E.~B. Davies and J.~T. Lewis,
\newblock {Commun.\ Math.\ Phys.} {\bf 17} (1970), 239.

\bibitem{84QC}
M. Ozawa,
\newblock {J. Math.\ Phys.} {\bf 25} (1984), 79.

\bibitem{85CA}
M. Ozawa,
\newblock {Publ.\ Res.\ Inst.\ Math.\ Sci., Kyoto Univ.} 
{\bf 21} (1985), 279.

\bibitem{IUO90}
N. Imoto, M. Ueda, and T. Ogawa,
\newblock {Phys.\ Rev.\ A} {\bf 41} (1990), 4127.

\bibitem{Yue83}
H.~P. Yuen,
\newblock {Phys.\ Rev.\ Lett.} {\bf 51} (1983), 719.

\bibitem{88MS}
M. Ozawa,
\newblock {Phys.\ Rev.\ Lett.} {\bf 60} (1988), 385.

\bibitem{Mad88}
J. Maddox,
\newblock {Nature} {\bf 331} (1988), 559.

\bibitem{89RS}
M. Ozawa,
\newblock in {\it Squeezed and Nonclassical Light}, 
edited by P. Tombesi and E.~R. Pike (Plenum, New York, 1989), 
pp.~263--286.

\bibitem{BV74}
V. B. Braginsky and Yu. I. Vorontsov, 
Uspehi Fiz.\ Nauk {\bf 114} (1974), 41 
[Sov.\ Phys.\ Usp.\ {\bf 17} (1975), 644].

\bibitem{CTDSZ80}
C.~M. Caves, K.~S. Thorne, R.~W.~P. Drever, V.~D. Sandberg, and M. Zimmermann,
\newblock {Rev.\ Mod.\ Phys.} {\bf 52} (1980), 341.

\bibitem{Cav85}
C.~M. Caves,
\newblock {Phys.\ Rev.\ Lett.} {\bf 54} (1985), 2465.

\bibitem{BDSW96}
C.~H. Bennet, D.~P. DiVincenzo, J.~A. Smolin, and W.~K. Wootters,
\newblock {Phys.\ Rev.\ A} {\bf 54} (1996), 3824.

\bibitem{VPRK97}
V. Vedral, M.~B. Plenio, M.~A. Rippin, and P.~L. Knight,
\newblock {Phys.\ Rev.\ Lett.} {\bf 78} (1997), 2275.

\bibitem{Dav76}
E.~B. Davies,
\newblock {\it Quantum Theory of Open Systems}
\newblock (Academic Press, London, 1976).

\bibitem{HK64}
R. Haag and D. Kastler,
\newblock {J. Math.\ Phys.} {\bf 5} (1964), 848.

\bibitem{Lud67}
G. Ludwig,
\newblock {Commun.\ Math.\ Phys.} {\bf 4} (1967), {331}.

\bibitem{Hel76}
C.~W. Helstrom,
\newblock {\it Quantum Detection and Estimation Theory}
\newblock (Academic Press, New York, 1976).

\bibitem{Hol82}
A.~S. Holevo,
\newblock {\it Probabilistic and Statistical Aspects of Quantum Theory}
\newblock (North-Holland, Amsterdam, 1982).

\bibitem{83CR}
M. Ozawa,
\newblock in {\it Probability Theory and Mathematical Statistics}, 
edited by K. It\^{o} and J.~V. Prohorov, 
Lecture Notes in Mathematics {\bf 1021}
(Springer, Berlin, 1983), pp.~518--525, 

\bibitem{Sch60a}
R. Schatten,
\newblock {\it Norm Ideals of Completely Continuous Operators}
\newblock (Springer, New York, 1960).

%\bibitem{Kad52}
%R.~V. Kadison,
%\newblock {\it Ann.\ of Math.} {\bf 56} (1952), 494.

\bibitem{BR79}
O. Bratteli and D.~W. Robinson,
\newblock {\it Operator Algebras and Quantum Statistical Mechanics I}
\newblock (Springer, New York, 1979).

\bibitem{97OQ}
M. Ozawa,
\newblock {Ann.\ Phys.\ (N.Y.)} {\bf 259} (1997), 121.

\bibitem{97QQ}
M. Ozawa,
\newblock in {\it Quantum Communication, Computing, and Measurement},
edited by O. Hirota, A.~S. Holevo, and C.~M. Caves 
(Plenum, New York, 1997), pp.~233--241.

\bibitem{98QS}
M. Ozawa,
\newblock {Fortschr.\ Phys.} {\bf 46} (1998), 615.

\bibitem{Kra83}
K. Kraus,
\newblock {\it States, Effects, and Operations: Fundamental Notions of Quantum
  Theory},
\newblock {Lecture Notes in Physics {\bf 190}} (Springer, Berlin, 1983).

\bibitem{GL66}
J.~P. Gordon and W.~H. Louisell,
\newblock in  {\it Physics of Quantum Electronics}, 
edited by J.~L. {Kelly, Jr.}, B. Lax, and P.~E. Tannenwald 
(McGraw-Hill, New York, 1966), pp.~833--840. 

\bibitem{85CC}
M. Ozawa,
\newblock {J. Math.\ Phys.} {\bf 26} (1985), 1948.

\bibitem{Cho72}
M.~D. Choi,
\newblock {Can.\ J. Math.} {\bf 24} (1972), 520.

\bibitem{Kad65}
R.~V. Kadison,
\newblock {Topology} {\bf 3} (1965), 177.

\end{references}
\end{document}